\documentclass[11pt] {article}
\usepackage{graphicx}

\usepackage{amssymb}
\usepackage{amsmath}
\usepackage{appendix}
\input psfig.sty

\newtheorem{THEO}{Theorem}
\newtheorem{ALGo}[THEO]{Algorithm}
\newtheorem{ASSU}[THEO]{Assumption}
\newtheorem{CONJ}[THEO]{Conjecture}
\newtheorem{CORO}[THEO]{Corollary}
\newtheorem{DEFI}[THEO]{Definition}
\newtheorem{EXAM}[THEO]{Example}
\newtheorem{FACT}[THEO]{Fact}
\newtheorem{LEMM}[THEO]{Lemma}
\newtheorem{PROB}[THEO]{Problem}
\newtheorem{PROP}[THEO]{Proposition}
\newtheorem{REMA}[THEO]{Remark}
\newcommand{\theo}{\begin{THEO}}
\newcommand{\algo}{\begin{ALGo} \rm}
\newcommand{\assu}{\begin{ASSU} It}
\newcommand{\conj}{\begin{CONJ}}
\newcommand{\coro}{\begin{CORO}}
\newcommand{\defi}{\begin{DEFI}}
\newcommand{\exam}{\begin{EXAM} \rm}
\newcommand{\fact}{\begin{FACT}}
\newcommand{\lemm}{\begin{LEMM}}
\newcommand{\prob}{\begin{PROB} \rm}
\newcommand{\prop}{\begin{PROP}}
\newcommand{\rema}{\begin{REMA} \rm}
\newcommand{\etheo}{\end{THEO}}
\newcommand{\ealgo}{\end{ALGo}}
\newcommand{\eassu}{\end{ASSU}}
\newcommand{\econj}{\end{CONJ}}
\newcommand{\ecoro}{\end{CORO}}
\newcommand{\edefi}{\end{DEFI}}
\newcommand{\eexam}{\end{EXAM}}
\newcommand{\efact}{\end{FACT}}
\newcommand{\elemm}{\end{LEMM}}
\newcommand{\eprob}{\end{PROB}}
\newcommand{\eprop}{\end{PROP}}
\newcommand{\erema}{\end{REMA}}
\newcommand{\dt}{\mbox{\LARGE{.}}}

\def\z{\phantom{0}}

\newcommand{\qed}{\Box}

\def\Y{\mbox{\boldmath $Y$}}

\def\c{\mbox{\boldmath $c$}}

\def\d{\mbox{\boldmath $d$}}
\def\e{\mbox{\boldmath $e$}}

\def\p{\mbox{\boldmath $p$}}
\def\q{\mbox{\boldmath $q$}}
\def\r{\mbox{\boldmath $r$}}

\def\x{\mbox{\boldmath $x$}}
\def\y{\mbox{\boldmath $y$}}
\def\z{\mbox{\boldmath $z$}}

\def\0{\mbox{\boldmath $0$}}
\def\beqa{\begingroup \arraycolsep 1pt\begin{eqnarray}}
\def\eeqa{\end{eqnarray}\endgroup}
\def\beqas{\begingroup \arraycolsep 1pt\begin{eqnarray*}}
\def\eeqas{\end{eqnarray*}\endgroup}

\def\z{\phantom{0}}

\newcommand{\supt}{^\intercal}

\newcommand{\bdot}{\kern -.5pt\raise .2ex\hbox{\large\bf .}\kern -.3pt}

\newcommand{\czero}{{\tbi C}\raisebox{-3pt}{\scriptsize\bf 0}}

\newcommand{\pzero}{{\tbi P}\raisebox{-3pt}{\scriptsize\bf 0}}

\def\ppad{\cal{PPAD}}
\def\lcp{{LCP}}
\def\elcp{{ELCP}}
\def\llcp{{LLCP}}
\def\sg{SG}

\def\usr{\boldsymbol{\mathbf{USR}}}

\newcommand{\sol}{{SOL}}
\newcommand{\sne}{SNE}                          
\newcommand{\feas}{{FEA}}
\newcommand{\sr}{SR}
\newcommand{\sd}{SD}

\newcommand{\sz}[1]{\hbox{\tbi #1\lower3pt\hbox{\scriptsize\bf 0}}}

\def\r22{\boldsymbol{\mathcal{R}^{2 \times 2}}}
\def\rnn{\boldsymbol{\mathcal{R}^{n \times n}}}
\def\rmm{\boldsymbol{\mathcal{R}^{m \times m}}}
\def\rmn{\boldsymbol{\mathcal{R}^{m \times n}}}

\def\rmnp{\boldsymbol{\mathcal{R}^{m \times n}_{+}}}
\def\rmnpp{\boldsymbol{\mathcal{R}^{m \times n}_{++}}}
\def\rmpp{\boldsymbol{\mathcal{R}^{m}_{++}}}

\def\rn1{\boldsymbol{\mathcal{R}^{n}}}
\def\rm1{\boldsymbol{\mathcal{R}^{m}}}

\def\rmonepp{\boldsymbol{\mathcal{R}^{m}_{++}}}
\def\rmonep{\boldsymbol{\mathcal{R}^{m}_{+}}}
\def\rmone{\boldsymbol{\mathcal{R}^{m}}}

\def\rone{\boldsymbol{\mathcal{R}}}
\def\ronep{\boldsymbol{\mathcal{R}_{+}}}
\def\ronepp{\boldsymbol{\mathcal{R}_{++}}}

\def\qn1{\boldsymbol{\mathcal{Q}^{n \times 1}}}
\def\rsu{\boldsymbol{\mathbf{RSU}}}
\def\csu{\boldsymbol{\mathbf{CSU}}}
\def\su{\boldsymbol{\mathbf{SU}}}

\def\czero{\boldsymbol{\mathbf{C_0}}}
\def\c{\boldsymbol{\mathbf{C}}}
\def\cplus{\boldsymbol{\mathbf{C^+_0}}}
\def\cstar{\boldsymbol{\mathbf{C^*_0}}}

\def\rzero{\boldsymbol{\mathbf{R_0}}}

\def\r{\boldsymbol{\mathbf{R}}}

\def\q{\boldsymbol{\mathbf{Q}}}

\def\pzero{\boldsymbol{\mathbf{P_0}}}
\def\qzero{\boldsymbol{\mathbf{Q_0}}}

\def\estar{\boldsymbol{\mathbf{E^*}}}

\def\l{\boldsymbol{\mathbf{L}}}

\def\e{\boldsymbol{\mathbf{E}}}
\def\ezero{\boldsymbol{\mathbf{E_0}}}

\def\p{\boldsymbol{\mathbf{P}}}
\def\czero{\boldsymbol{\mathbf{C_0}}}

\def\y{\boldsymbol{\mathbf{Y}}}
\def\x{\boldsymbol{\mathbf{X}}}

\def\np{\cal{NP}}

\newcommand{\bq}{\begin{quote}}
\newcommand{\eq}{\end{quote}}

\begin{document}
\title{\bf A direct reduction of PPAD Lemke-verified  linear complementarity problems to bimatrix games}
\author{
Ilan Adler
\thanks{Department of Industrial Engineering and Operations
Research, University of California, Berkeley, CA 94720.}
\quad \quad
Sushil Verma
\thanks{SignalDemand, Inc.,
340 Brannan St, 4th Floor, San Francisco, CA 94107.}
}

\bibliographystyle{plain}

\pagestyle{myheadings}


\date{January 2013}

\maketitle
\begin{abstract}

\medskip

  The linear complementarity problem, $\lcp(q,M)$, is defined as follows. For given $ M \in \rmm, q \in \rmone$,  find $z$ such that
  $q + M z  \geq  0, \; z \geq 0,\;      z\supt (q + M z) =  0$, or  certify that there is no such $z$.
  It is well known that the problem of finding a Nash equilibrium for a bimatrix game (2-NASH) can be formulated as a  linear complementarity problem (LCP). In addition,  2-NASH is known to be complete in the complexity class PPAD (Polynomial-time Parity Argument Directed).
  However, the ingeniously constructed reduction (which is designed for any PPAD problem) is very complicated, so while of great theoretical significance, it is not practical for actually solving an LCP via 2-NASH, and it may not provide  the potential insight that can be gained from studying the game obtained from a problem  formulated as an LCP (e.g. market equilibrium). The main goal of this paper  is the construction of a simple explicit reduction of any $\lcp(q,M)$ that can be verified as belonging to $\ppad$
  via the graph induced by the generic Lemke algorithm with some positive covering vector $d$, to a symmetric 2-NASH. In particular, any endpoint of this graph (with the exception of the initial point of the algorithm) corresponds to either a solution or to a so-called secondary ray. Thus, an LCP problem is verified as belonging to $\ppad$ if any secondary ray can be used to construct, in polynomial time, a certificate that there is no solution to the problem. We achieve our goal by showing that for any $M,q$ and positive $d$ satisfying a certain nondegeneracy assumption with respect to $M$, we can simply and directly construct a symmetric 2-NASH whose Nash equilibria correspond one-to-one to the end points of the graph induced by $\lcp(q,M)$ and the Lemke algorithm with a covering vector $d$. We note that for a given $M$ the reduction works for all positive $d$ with the exception of a subset of measure $0$.

\end{abstract}
\section{Introduction}\label{sec:intro}
  The linear complementarity problem $\lcp(q,M)$ is defined as
\[ \mbox{ For given } q \in \rmone,\; M \in \rmm,\;\;
\mbox{find $z \in \rmone $ such that} \;\; q + M z  \geq  0, \; z \geq 0,\;      z\supt (q + M z) =  0.\]
  The LCP is notable for its wide range of applications,
  from well understood and relatively easy to solve problems, such as
  linear and convex quadratic programming problems, to $\np$-hard
  problems.
 A major effort in LCP theory had been the study of variants of Lemke's
 algorithm, a Simplex-like vertex following algorithm.
  In particular, for a given positive {\it covering} vector $d$, the Lemke($d$) algorithm
  goes through a path of adjacent vertices of the `extended' $\lcp(q,M)$  (denoted by $\elcp(d,q,M)$) where $d$
  is attached to $M$ with an artificial variable $z_0$. Assuming (without loss of generality) that
  $\elcp(d,q,M)$ is nondegenerate,  Lemke($d$) is guaranteed to terminate in a finite number of steps
  with either a solution to the original problem or with a {\it secondary ray} of $\elcp(d,q,M)$. If the secondary rays
  can certify (in polynomial time) that there is no solution to $\lcp(q,M)$, we say that the problem
  is {\it Lemke($d$)-resolvable}.
One of the major themes of LCP research over the years has been the search for classes of matrices $M$ and covering vectors $d$
  for which $\lcp(q,M)$ is Lemke($d$)-resolvable for all $q$.
   Several such classes (usually applicable for all $d>0$) were identified (see e.g. \cite{cps92},
\cite{m88} and the references therein).

\medskip

The introduction of the $\ppad$ ({\it Polynomial-time Parity Argument Directed}) complexity
  class in \cite{p94}  provides an effective and elegant framework for analyzing the complexity
  of Lemke($d$)-resolvable linear complementarity problems since, in general,
  the directed graph induced by the Lemke($d$) algorithm for a given $\lcp(q,M)$ can be used to verify the membership of the problem in  $\ppad$.
  We say in this case that the problem is {\it Lemke($d$) $\ppad$-verified}.
  This development is
  significant with respect to LCP theory since it has been shown in \cite{mp91} that
  if $\ppad$ is $\np$-hard then $\np = \mathrm{Co}\np$,
   lending support to the long standing informal
  belief that LCPs resolvable
  by Lemke($d$) algorithm are in some way special.

  What makes the class $\ppad$ particularly
  interesting  is the fact that several
  well known problems, such as finding a
 Brouwer fixed-point, were identified in \cite{p94} as
  $\ppad$-complete. The discovery, in a string of papers
  (\cite{dp05}, \cite{dp05a}, \cite{cd05} and \cite{cd05a}),
  that finding a Nash equilibrium
   of a bimatrix game (2-NASH) is $\ppad$-complete has significant
   consequences in the context of LCP theory. It has been known since the early days of LCP research that the 2-NASH problem can be formulated as an LCP with roughly the same size and with the coefficient matrix belonging to one of several well known classes resolvable by Lemke($d$) algorithm. The fact that
   2-NASH is $\ppad$-complete  means that any $\lcp(q,M)$ verifiable as a member in $\ppad$ (including all classes that contain 2-NASH) can be reduced to a 2-NASH problem. However, the known reduction is quite complicated. It requires  several stages that involve reducing the given
$\lcp(q,M)$ to finding an approximate Brouwer fixed point of an appropriate function, followed by reducing the
latter to 3-graphical NASH (using small polymatrix games to simulate the computation of certain simple
arithmetic operations), and finally, reducing the 3-graphical NASH to 2-NASH\footnote{A clear `bird's-eye view' description of the reduction can be found in \cite{dgp09}.}.
     While there seems to be no discussion in the vast literature on LCP suggesting the possibility
    that Lemke($d$) $\ppad$-verified LCPs
   can be reduced to 2-NASH, the discovery
   that 2-NASH is $\ppad$-complete motivated us to search for the existence of a direct simple reduction of such problems to 2-NASH.

The main result of this paper is the introduction of  a direct, simple reduction of almost any Lemke($d$) $\ppad$-verified linear complementarity problems
to a symmetric 2-NASH.
 In fact, we introduce a stronger result as  follows.  Consider a generic Lemke($d$) $\lcp(q,M)$
 (which we call $\llcp(d,q,M)$) whose `solutions' are defined to be either actual solutions of $\lcp(q,M)$ or secondary rays of $\elcp(d,q,M)$.
  Obviously this problem\footnote{Where, as we assume without loss of generality, its extended form, $\elcp(d,q,M)$, is nondegenerate.} is  Lemke($d$) $\ppad$-verified. Through a series of steps we show how to construct a symmetric bimatrix game whose equilibria correspond one-to-one to the `solutions' of
  $\llcp(d,q,M)$. The point is that if $\lcp(q,M)$ is Lemke($d$) $\ppad$-verified, the
Nash equilibria of the constructed bimatrix game correspond to solutions (or certifications for infeasibility) of the given problem, which means that the constructed bimatrix game properly resolves an $\lcp(q,M)$ if it is Lemke($d$)-verified.

We begin by reviewing the  Lemke($d$) algorithm (in
Section \ref{sec:lcp}) and bimatrix games  (in Section \ref{sec:nash}).  Next, we introduce (in Section \ref{sec:ppad})
the complexity class $\ppad$, and briefly discuss the notion of Lemke($d$) $\ppad$-verified LCPs. In addition, we present the majority of matrix classes known to be  Lemke($d$) $\ppad$-verified,
and identify a number of matrix classes whose corresponding LCPs are $\ppad$-complete.
We conclude the introductory sections by introducing  (in Section \ref{sec:llcp}) $\llcp(d,q,M)$,  the generic Lemke($d$) $\lcp(q,M)$.

  Our main results are presented in sections \ref{sec:re}-\ref{sec:nd}.
 We start by introducing in Section \ref{sec:re} a very simple reduction of $\lcp(q,M)$, where $M$ belongs to a class of matrices for
  which a solution is guaranteed to exist for all $q$, to a symmetric 2-NASH.
 The cost matrix of the resulting bimatrix game is composed of $M$
 with an extra row and column. In particular, we show that  the solutions of the given $\lcp(q,M)$ correspond one-to-one to the Nash equilibria which  use with positive probability for the pure strategy corresponding to the extra column of the cost matrix. Moreover, we show that the Nash equilibria which  do not use  the pure strategy corresponding to the extra column of the cost matrix of the resulting game correspond one-to-one to the so-called `secondary directions' of $\elcp(e,q,M)$. Note that at this stage we address only $e$ - the vector of all ones - as a
 covering vector, and that we do not reach yet our goal as the reduction may produce secondary {\it directions} rather than secondary {\it rays}.\footnote{A ray of
 $\elcp(d,q,M)$ is an unbounded edge of $\elcp(d,q,M)$ which includes its endpoint  (a vertex of $\elcp(q,M)$) together with a direction vector.}

 In Section \ref{sec:r_0}, we  extend the basic reduction above (by considering an augmented problem) so that the constructed bimatrix game produces either a solution for $\lcp(q,M)$, a secondary ray for $\elcp(e,q,M)$, or a non-zero vector which is a solution to $\lcp(e,M)$ (and is actually also
   a special case of a secondary direction of $\elcp(e,q,M)$).

   In Section \ref{sec:nd}, we show that if a secondary direction generated by the bimatrix game  constructed in the previous section is a nondegenerate solution of $\lcp(e,M)$, we can use it to compute, in strongly polynomial time either a solution for $\lcp(q,M)$ or a secondary ray for $\elcp(e,q,M)$; thus showing that the constructed bimatrix game indeed provides a `solution' for $\llcp(e,q,M)$.

 In Section \ref{sec:ext} we extend the results of the previous section  to accommodate the reduction of any  $\llcp(d,q,M)$ for which $d>0$ and
 $\lcp(d,M)$ is nondegenerate; thereby achieving our goal of showing that  any $\lcp(q,M)$ which is Lemke($d$) $\ppad$-verified (satisfying our nondegeneracy assumption as stated above) can be reduced to a symmetric 2-NASH.  We note that for any given $M$ and $q$, the reduction is workable for all positive covering vectors with the exception of a finite number of sets of measure $0$.

  The constructed reduction is particulary useful since it provides a bijection between the reducible LCPs and their corresponding 2-NASH problems. In particular, the simplicity of the reduction and its bijection property  allows for the practical use of the results of the extensive
  research on  `non Lemke type' 2-NASH algorithms for solving (or enumerating the solutions of) reducible
    LCPs . In addition, these reductions can be applied to investigate properties of solutions of reducible LCP via known properties of the associated 2-NASH problems.
We discuss these subjects together with additional concluding remarks
    in Section \ref{sec:remarks}.

Throughout the paper we denote by $e$ vectors all of whose entries are 1. Given a matrix $A$, we denote by $A_{i \dt}$
the i-th row of $A$, by $A_{\dt j}$ the j-th column of $A$, and by $A_{ij}$ the ij-th entry  of $A$. We denote by $\rmn, \rmnp$, and $\rmnpp$ the space of $m \times n$ real matrices,
the space of nonnegative $m \times n$ real matrices, and the space of positive $m \times n$ real matrices, respectively. Whenever $n=1$ we abbreviate
$\rmn$ to $\rmone$, and whenever $m=n=1$ we abbreviate $\rmn$ to $\rone$.
\section{LCP
and Lemke's algorithm}
\label{sec:lcp}
Given  $M \in \rmm,\; q \in \rm1,$ the {\it linear complementarity problem}, $\lcp(q,M)$, is defined as
\begin{subequations}
 \label{eq:lcp}
  \begin{eqnarray}
\mbox{find }   z \in \rm1\ \mbox{ such that } && \nonumber  \\
 \label{eq:lcp_1}
     q + M z   \geq  0, \;\; z \geq 0,&&\\ %
        \label{eq:lcp_2}
        z\supt (q + M z) =  0. &&
  \end{eqnarray}
\end{subequations}
Note that (\ref{eq:lcp_1})--(\ref{eq:lcp_2}) imply
 \begin{equation}
 \tag{1c}
   \label{eq:lcp_3}
 z_i( q_i + M_{ i \dt} z)=0,\;\
      i=1, \ldots, m.
   \end{equation}

We denote by $\feas(q,M)$  the set of all $z$ satisfying (\ref{eq:lcp_1}), and by
$\sol(q,M)$  the set of all $z$ satisfying (\ref{eq:lcp_1}) and (\ref{eq:lcp_2}).

 In this section we present
 the generic Lemke algorithm
 (the so-called Scheme I - see \cite{cps92}, 4.4.5).
 Given $\lcp(q,M)$ we define its {\it extended} version, with a {\it covering vector } $d>0$
 as
 \begin{equation}
   \elcp(d,q,M) \triangleq \{  z_0 \in \ronep,\; z \in \rmonep\; \:|\; q + d z_0  +  Mz  \geq 0
   \; \mbox{ and }\; z\supt (q + d z_0 + Mz)=0 \}. \nonumber
  \end{equation}
  Note that $\elcp(d,q,M)$ is composed of a polyhedral set intersected with one nonlinear complementarity constraint.
  Throughout the paper whenever we refer to vertices, edges and rays of
  $\elcp(e,q,M)$ we mean the vertices, edges and rays of
  the polyhedral set  associated with $\elcp(e,q,M)$.
      We assume that  $\elcp(d,q,M)$ is nondegenerate, that is that the polyhedral set associated with it
      is nondegenerate\footnote{There is no loss of generality in this assumption since
       if it is not satisfied, we perturb $q$
    by applying standard linear programming techniques.}.
         Let $(\bar{z}_0, \bar{z})  \in \elcp(d,q,M), \bar{w}= q + d \bar{z}_0 +    M\bar{z}$,
    and let $k$ be the
    the number of positive entries in $(\bar{z}_0, \bar{z},\bar{w})$.
    By the nondegeneracy assumption, $k$ is equal to either $m$ (in which case
    $(\bar{z}_0, \bar{z})$ is
    a vertex of $\elcp(d,q,M)$), or $m+1$ (in which case it is
    a point on an edge  of $\elcp(d,q,M)$). If  a vertex of  $\elcp(d,q,M)$  is contained in an edge,
    we say that the vertex is an {\it endpoint}
    of the edge. If  an edge of $\elcp(d,q,M)$ is unbounded then it corresponds to a ray
    of  $\elcp(d,q,M)$, which can be presented as
 \begin{equation}
  \label{eq:ray}
    \left\{\left[
\begin{array}{c}
z_0 \\
 z\\
\end{array}
\right]
  |
 \left[
 \begin{array}{c}
z_0 \\
z \\
\end{array}
\right]
    =
    \left[
 \begin{array}{c}
 \bar{z}_0\\
   \bar{z}\\
\end{array}
\right]
      +
      \left[
 \begin{array}{c}
 \bar{u}_0\\
 \bar{u}\\
\end{array}
\right]
\lambda\
   \mbox{ for all } \lambda \geq 0 \right\}
\end{equation}

where
   \begin{subequations}
    \label{eq:sr}
    \begin{eqnarray*}
   (\bar{z}_0, \bar{z})
     \mbox{ is a vertex of } \elcp(d,q,M),  \hspace{53mm} \\
  \bar{u} \in \sol(d\bar{u}_0,M),\; \bar{u}_0 \in \{0,1\}\;
   \mbox{ and }
   (\bar{u}_0, \bar{u}) \neq 0,
    \hspace{34mm}\\
   \bar{z}\supt (d \bar{u}_0 + M \bar{u})=0\;
   \mbox{ and }  \;\bar{u}\supt (q + d \bar{z}_0 + M \bar{z})=0.
    \hspace{35.3mm}
  \end{eqnarray*}
  \end{subequations}

    Consider the  ray of $\elcp(d,q,M)$ with
    $\bar{z}=0,\;\bar{z}_0=-\min_{1 \leq i \leq m} q_i,\;
      \bar{u}_0=1,\; \bar{u}=0.$
    We call this ray the {\it primary ray}, and its corresponding endpoint vertex the {\it initial vertex}.
   Any other  ray of $\elcp(d,q,M)$, can be characterized as
    \begin{subequations}
    \label{eq:sr}
    \begin{eqnarray}
     \label{eq:sr_1}
   (\bar{z}_0, \bar{z})
     \mbox{ is a vertex of } \elcp(d,q,M),   \hspace{53mm}  \\
  \label{eq:sr_2}
    \bar{u} \in \sol(d\bar{u}_0,M)\setminus\{0\},\; \bar{u}_0 \in \{0,1\}\;
   \mbox{ and }
   \;e\supt \bar{u}=1  \mbox{ whenever } \bar{u}_0=0,\\
    \label{eq:sr_3}
   \bar{z}\supt (d \bar{u}_0 + M \bar{u})=0\;
   \mbox{ and }  \;\bar{u}\supt (q + d \bar{z}_0 + M \bar{z})=0.
    \hspace{35.3mm}
  \end{eqnarray}
  \end{subequations}
     We call such a ray, a {\it secondary ray}.
     We denote the set of all secondary rays of $\elcp(d,q,M)$ by
      \[
    \sr(d,q,M) \triangleq \{
   (\bar{z}_0,\bar{z}, \bar{u}_0,\bar{u})  \; \mbox{ satisfying \eqref{eq:sr_1}--\eqref{eq:sr_3}}\}.
 \]
    Note that a secondary ray $(\bar{z}_0,\bar{z}, \bar{u}_0,\bar{u})$ has two components, a vertex
   $(\bar{z}_0,\bar{z})$ of $\elcp(d,q,M)$ and what we call a {\it secondary direction} $(\bar{u}_0,\bar{u})$ as defined in \eqref{eq:sr_2}.
   Specifically, we denote the set of all secondary directions of $\elcp(d,q,M)$ as
       \[
    \sd(d,M) \triangleq \{
   (\bar{u}_0,\bar{u})  \; \mbox{ satisfying \eqref{eq:sr_2}}\}.
 \]
     We distinguish between two types of secondary directions (and rays), according to whether
     $\bar{u}_0=0$, which we call a {\it type $0$ secondary direction}, or
     $\bar{u}_0=1$, which we call a {\it type $1$ secondary direction}.
    Specifically,  for $k=0,1$, we denote the set of all type $k$ secondary directions of $\elcp(d,q,M)$ by
      \begin{eqnarray*}
   \sd_k(d,M) \triangleq \{
   (\bar{u}_0,\bar{u}) \in \sd(d,q,M)\;|\;  \bar{u}_0=k \}.
  \end{eqnarray*}
    Similarly,  for $k=0,1$, we we denote the set of all type $k$ secondary rays of $\elcp(d,q,M)$ by
    \begin{eqnarray*}
   \sr_k(d,q,M) \triangleq \{
   (\bar{z}_0,\bar{z}, \bar{u}_0,\bar{u}) \in \sr(d,q,M)\;|\;  \bar{u}_0=k \}.
     \end{eqnarray*}
 Starting with the initial vertex of $\elcp(d,q,M)$, the generic Lemke($d$) algorithm
    traces a unique\footnote{The uniqueness is due to the assumption that $\elcp(d,q,M)$ is nondegenerate.}
finite path of adjacent  vertices of $\elcp(d,q,M)$, terminating with either a solution to
$\lcp(q,M)$ or with a secondary ray of $\elcp(d,q,M)$. Specifically, the algorithm ends  with either a vertex
$(\bar{z}_0,\bar{z})$ of $\elcp(d,q,M)$
with $\bar{z}_0=0$
(so $\bar{z} \in \sol(q,M)$),
 or  with a secondary ray
   $(\bar{z}_0,\bar{z},\bar{u}_0,\bar{u})\in  \sr(d,q,M)$ with $\bar{z}_0>0$.
  We say that Lemke(d) {\it resolves} a given $\lcp(q,M)$ if either it ends with
$\bar{z} \in \sol(q,M)$, or if the terminal secondary ray can certify that $\sol(q,M) = \emptyset$.
Whenever Lemke($d$) resolves $\lcp(q,M)$ we say that $\lcp(q,M)$ is {\it Lemke($d$)-resolvable}.

 Ever since the introduction of the Lemke algorithm \cite{l65}, extensive research efforts focused on identifying classes of
matrices $M$ for which  $\lcp(q,M)$ is Lemke($d$)-resolvable for all $q$. In the following we discuss two major groups
of matrices containing almost all known classes of matrices $M$ for which $\lcp(d,q,M)$ is Lemke($d$)-resolvable for all $q$.

The first group is based on the idea that if
$\sr(d,q,M) = \emptyset$, then Lemke($d$) outputs
$\bar{z} \in \sol(q,M)$. Specifically, we consider the class of  $d$-regular matrices
(see \cite{cps92}, 3.9.20)
as defined below.

\noindent{\bf Definition} \hspace{2mm}
   Given $M \in \rmm$ and $d \in \rmpp$, we say that $M$ is {\it $d$-regular} if
   $\sol(d \tau ,M)=\{0\}$ for all $\tau \in \ronep $. We denote the class of $d$-regular matrices
     by $\r(d)$.

     It follows that if $M \in \r(d)$, then for all $q$, $\elcp(d,q,M)$ has no secondary directions
     and thus no secondary rays. That is,
        $\sr(d,q,M) = \emptyset$ for  all $q$.
          Recalling that   Lemke($d$)  terminates in a finite number of steps with
     either $\bar{z} \in \sol(q,M)$ or with
        $(\bar{z}_0,\bar{z},\bar{u}_0,\bar{u}) \in  \sr(d,q,M)$,
      we  conclude that whenever $M \in \r(d)$, $\lcp(q,M)$ is  Lemke($d$)-resolvable for all $q$.

     \noindent
     {\bf Remark}\footnote{The definitions of all the matrix classes which are mentioned in this paper can be found in
     \cite{cps92}.} \hspace{1mm}
       It is well  known that $M$ belongs to the {\it strictly semimonotone}  matrix class (which is denoted by $\e$) if and only if $\sol(q,M) = \{0\}$ for all $q \geq 0$ (see \cite{cps92},3.9.11). Thus, it follows that for all $d > 0, \e \subset \r(d)$ .
         In addition, $\e$ properly  contains the {\it  strictly copositive} matrix class ($\c$),
 and the class of all matrices whose principle minors are positive ($\p$).
  Thus, we observe that $\lcp(q,M)$ with $M$ in $\e,\c$ or $\p$ is Lemke($d$)-resolvable for all $d>0$ and all $q$.

      The second group  includes classes of matrices for which
    $\sr(d,q,M) \neq \emptyset$ implies that $\sol(q,M)=\emptyset$.
    Specifically,
    most matrix classes with this property that have been identified in the LCP literature share the following  property:
   \vspace{-1.5mm}
   \begin{subequations}
   \label{eq:usr}
  \begin{eqnarray}
  \label{eq:usr_1}
     & &   \sd_1(d,M)=\emptyset\\
    \label{eq:usr_2}
     & & \sr_0(d,q,M)  \ne \emptyset\
      \Rightarrow   \feas(q,M) = \emptyset.
  \end{eqnarray}
\end{subequations}
 We denote by $\usr(d)$ (for {\it useful secondary rays}) the class of all matrices $M$ for which \eqref{eq:usr_1}-\eqref{eq:usr_2}
  are satisfied for all $q \geq 0$ .

    The class of matrices that satisfy  (\ref{eq:usr_1}) is defined below.

     \noindent{\bf Definition} \hspace{2mm}
   Given $M \in \rmm$ and $d \in \rmpp$, we say that $M \in \rmm$ is { \it $d$-semiregular} if
   $\sol(d ,M)=\{0\}$. We denote the class of $d$-semiregular matrices
     by $\rzero(d)$.

    \noindent{\bf Remarks} \hspace{2mm}
    \begin{enumerate}
    \item
     While the term `$d$-semiregular' is introduced here for the first time, the class itself has been
    introduced in \cite{g73} under the name $\estar(d)$.
    \item
     If $M \in \usr(d)$ then the existence of a secondary ray for $\elcp(d,q,M)$ implies
    that $\sol(q,M) = \emptyset$. Hence, any $\lcp(q,M)$ with $M \in \usr(d)$ is Lemke($d$)-resolvable for all $q$.
    \item
    $\r(d) \subset \usr(d)$.
    \item
     It is well  known that $M$ belongs to the {\it  semimonotone}  matrix class (which is denoted by $\ezero$) if and only if $\sol(q,M) = \{0\}$
        for all $q > 0$ (see \cite{cps92},3.9.3). Thus, it follows that $\ezero \subset \rzero(d)$ for all $d >0$.
         In addition, $\ezero$ properly  includes the {\it  copositive} matrix class ($\czero$),
 and the class of all matrices whose principle minors are nonnegative ($\pzero$).
\item
     There are two well known classes of matrices, $\l$ and $\qzero \cap \pzero$, which are known to be in
     $\usr (d)$ for all $d>0$.
         In particular, major matrix classes, including
      {\it Column Sufficient} ($\csu$), {\it Row Sufficient} ($\rsu$), and {\it  Sufficient} ($\su$),
        are subsets of  $\pzero \cap \qzero$, while {\it  Copositive Plus} ($\cplus$), and {\it  Copositive Star} ($\cstar$) are subsets of
        $\l$. Hence, $\lcp(q,M)$ where $M$ belongs to any of these classes of matrices is Lemke($d$)-resolvable.
        For a discussion of these and other Lemke($d$)-resolvable classes see \cite{cps92} and \cite{m88}.
        Figure 1 at the end of Section \ref{sec:llcp} depicts the relationship among these classes.
        \end{enumerate}
\section{Bimatrix Games}
 \label{sec:nash}
Let $A,B \in \rmn$  be the cost matrices of the row and column players of a bimatrix game.
   A {\it Nash equilibrium} of this game is a pair of vectors $x \in \rm1,\;y \in \rn1 $
   (representing mixed strategies for the row and column players respectively),
   satisfying
   \[
   A y  \geq e(x\supt A y), \; B\supt x \geq e(x\supt B y),\; e\supt x=e\supt y=1, \;x \geq 0,\; y \geq 0.
   \]
   To simplify the presentation we restrict our attention to {\it symmetric  bimatrix games}  where $A = B\supt$.    In particular, it has been shown in the seminal paper \cite{n51} that every symmetric bimatrix game has a symmetric Nash equilibrium (that is, a Nash equilibrium where $x=y$).
   In addition, it is well known that the Nash equilibria for any bimatrix game
   with cost matrices $A,B$ (which can be assumed, without loss of generality, to be positive)  can be easily
    extracted  from the symmetric equilibria of the symmetric bimatrix game with cost matrix
  $\left(
\begin{array}{cc}
 0 & A\\
 B\supt & 0
\end{array}
\right)
$.

  Given
   $C \in \rnn$, we denote by $\sg(C)$ the symmetric bimatrix game where the row and column players' cost matrix is $C$.
  %
 We say that $x \in \rn1$ is
a {\it symmetric  Nash equilibrium} of $\sg(C)$ if
\begin{subequations}
\label{eq:ne}
  \begin{eqnarray}
  \label{eq:ne_1}
     & &  C x \geq e (x\supt  C  x),\\
    \label{eq:ne_2}
     & & x \geq 0,  \\%
      \label{eq:ne_3}
     & &  e\supt x =1. 
  \end{eqnarray}
\end{subequations}
Note that since $x\supt C x= \sum_{i=1}^m x_i (C_{ i \dt} x)$, (\ref{eq:ne_1})--(\ref{eq:ne_2}) imply
 \begin{equation}
 \tag{9d}
   \label{eq:ne_4}
 x_i( C_{ i \dt} x -x\supt  C x )=0,\;\
      i=1, \ldots, n.
   \end{equation}
  We denote by $\sne(C)$ the set of symmetric Nash equilibria  of $\sg(C)$.
  We refer to the problem of finding a symmetric Nash equilibrium for $\sg(C)$ as
  {\it solving} $\sg(C)$.

  There are several ways of formulating the problem of finding a Nash equilibrium of a bimatrix game as a linear complementarity problem
  (\cite{cd68}, \cite{e71}, \cite{mz91}, \cite{s06}). Here we adopt the reduction in \cite{s06}, where the problem of computing a
symmetric Nash equilibrium of a symmetric bimatrix game is presented as a linear complementarity problem.
In particular, let $C$ be the cost matrix of a symmetric bimatrix game. Without loss of generality we can  assume (by adding a sufficiently large constant to all the entries of $C$) that $C>0$. Solving $\sg(C)$ with $C>0$ can be reduced to $\lcp(-e,C)$
as described in \cite{s06}, and presented in the following theorem.
 \vspace{-6mm}
 \noindent
\theo
 \label{theo:sne_to_lcp}
 \hphantom \newline \noindent
 Suppose $C >0$.
\begin{description}
\item[(i)]
 Let $z \in \sol(-e,C)$.
   Then, $ z \frac{1}{e\supt z}\z \in \sne(C)$.
 \item[(ii)]
 Let $x \in \sne(C)$.
  Then, $x \frac{1}{x\supt C x} \in \sol(-e,C)$.
\end{description}
\etheo

\noindent {\bf Proof.} \hspace{3mm} (i) and (ii) can be easily verified by substitution.
 $\hspace*{\fill} \qed$
         \section{The Complexity Class PPAD}
          \label{sec:ppad}
The class $\ppad$ ({\it Polynomial-time Parity Argument Directed}), which
was introduced in the seminal paper \cite{p94}, is a class of problems which can be
presented as follows.

  \noindent{\bf Definition} \hspace{2mm}
  Given a directed graph with every node having in-degree and
out-degree at most one described by a polynomial-time computable
function $f(v)$ that outputs the predecessor and successor of a node
$v$, and a node $s$ (which we call the {\it initial source node})
 with a successor but no predecessor, find a node $t \neq s$ which is either
  a {\it sink} (a node with no  successor) or  a {\it source} (a node with no predecessor), but not both.
   We call such a graph the {\it $\ppad$  graph associated with the problem}.

  Many important problems,
  such as the Brouwer fixed-point problem, the search versions of
  Smith's theorem, the Borsuk-Ulam theorem and, as previously discussed, Nash
  equilibrium of bimatrix game, belong to this class \cite{p94}.
Interestingly,
 the problems in $\ppad$ are generally believed   not to be $\np$-hard since it has been shown in
  \cite{mp91} that
      if there exists a  $\ppad$ problem which is $\np$-hard then
       $\np = \cal{C}$o$\cal{NP}$. What makes the study of this class attractive is that
        it has been shown that several
       problems within the class (such as the Brouwer fixed-point problem) are $\ppad$-complete
        with strong circumstantial evidence that these problems are not
       likely to have a polynomial time algorithm \cite{hpv89}.

      The $\ppad$ complexity class seems to be a natural framework for analyzing the computational complexity of
      Lemke($d$)-resolvable $\lcp(q,M)$, as the underlying graph of  Lemke($d$)  whose nodes correspond to
       the  vertices and  edges  of $\elcp(d,q,M)$
      has a structure reminiscent of a $\ppad$ graph. In particular, given $\elcp(d,q,M)$, we define its  associated graph
      (which we call the {\it Lemke($d$) graph associated with $\lcp(q,M)$}),
      as the directed graph $G(d,q,M)$ whose nodes correspond to the vertices and edges (including rays) of $\elcp(d,q,M)$. There is an arc
      $(u,v)$  of $G(d,q,M)$ if and only if either $u$ corresponds to a vertex of $\elcp(d,q,M)$, $v$ corresponds to an edge of $\elcp(d,q,M)$
      and the vertex corresponding to $u$ is the tail of the edge corresponding to $v$; or
      $u$ corresponds to an edge of $\elcp(d,q,M)$, $v$ corresponds to a vertex of $\elcp(d,q,M)$
      and the vertex corresponding to $v$ is the head of the edge corresponding to $v$.
       The orientations of the edges are determined according the scheme presented in \cite{t76}. We identify the node
      associated with the primary ray of $\elcp(d,q,M)$ as the required special source node of a $\ppad$ graph.
      Given (as we assume) that
      $\elcp(d,q,M)$ is nondegenerate, we have that every node of $G(d,q,M)$ is incident to at most two edges, and that there are no isolated
      nodes. Thus, $G(d,q,M)$ is a nonempty collection of simple directed paths. In addition, any node incident to only one other node (except for the node associated with the primary ray) corresponds to either a solution of  $\lcp(q,M)$ or to a secondary ray of  $\elcp(d,q,M)$.
      Thus, if for a given $\lcp(q,M)$ and a covering vector $d$, the secondary rays
      of $\elcp(d,q,M)$ can certify (in polynomial time in the size of $\lcp(q,M)$) that $\sol(q,M)=\emptyset$, we can conclude that
      $\lcp(q,M)$ is in $\ppad$. Whenever this is the case, we say that {\it $\lcp(q,M)$ is Lemke($d$) $\ppad$-verified}.

           Indeed, in \cite{p94},
      one of the first examples of a $\ppad$ problem is an $\lcp(q,M)$ where $M \in \p$.
      While it is customary in the literature of linear complementarity to discuss methods for solving  $\lcp(q,M)$ under the assumption that
      $M$ possesses some special properties, it creates difficulties from an algorithmic complexity point of view, as verifying these properties may be by itself a hard problem (e.g. identifying a $\p$ matrix is $\cal{C}$o$\cal{NP}$ complete \cite{c94}). Thus, in  \cite{p94},
      the problem at hand (which is called $\p-LCP$) is defined as follows. Given $q,M$, either find $z \in \sol(q,M)$,
      or provide a certificate (with size polynomial in the size of the problem) for $M \not \in \p$.
      Motivated by the discussion in \cite{p94} we consider the following generic problem.

      \noindent
  $\y-\lcp(q,M)$\;:\; Given $M \in \rmm,\; q \in \rm1$ and a matrix class $\y$, find
 one of the following:

   (1) $z \in \sol(q,M),\;$  (2) a certificate that $\sol(q,M)=\emptyset,\;$  (3) a certificate that $M \not \in \y$.

     \medskip

     We say that $\y-\lcp(q,M)$ is {\it Lemke($d$)-\ppad-verified} if any $(\bar{z}_0,\bar{z},\bar{u}_0,\bar{u}) \in \sr(d,q,M)$ leads
     (in  polynomial time in the size of $\elcp(d,q,M)$) to (2) or (3) above.


         \noindent{\bf Remark} \hspace{2mm}
      Following the discussion in section \ref{sec:lcp}, we have that
       $\usr(q,M)-\lcp(q,M)$ is Lemke($d$) $\ppad$-verified and for all $q$ and $d>0$.
     In particular, considering
     the remarks at the end of Section \ref{sec:lcp}, we can conclude that
     $\l \cup (\pzero \cap \qzero)-\lcp(q,M)$ is Lemke($d$) $\ppad$-verified and for all $q$ and $d>0$.
         \section{The generic Lemke($d$) linear complementarity problem }
\label{sec:llcp}
     As stated in the introduction, it has been established that the problem of finding a Nash equilibrium for a bimatrix game  is
     $\ppad$-complete. Moreover, since  solving any bimatrix game is  polynomially reducible to solving a symmetric bimatrix game,
      we have that the problem of finding a
     symmetric Nash equilibrium for a symmetric bimatrix game, as presented in Section \ref{sec:nash}, is also $\ppad$-complete. In particular, it is shown there that this problem can be represented as an $\lcp(-e,M)$ where $M>0$. Since $M>0$ implies that $M \in \c$
     (the class of all matrices
     for which $0 \neq x \in \rmonep$ implies that $x \supt M x >0$), and considering the remark at the end of the previous section,
       we conclude that $\c-\lcp$ is $\ppad$-complete as well. In Figure 1, we display
     the relationship among the classes of matrices discussed in  previous  sections. An arrow from a class $\x$ to a class $\y$ indicates that
     $\x \subset \y$. So for any class $\y$ reachable by a directed path from class $\c$  in Figure 1
      we have that if $\y-\lcp$ is in $\ppad$ then it is $\ppad$-complete.
     Note that the class $\usr(d)$ (for any $d>0$) contains all the classes of matrices $\y$ identified in the previous section as a classes for which
     $\y-\lcp(q,M)$ is Lemke($d$) $\ppad$-verified and for all $q$ and $d>0$.

     \medskip

       \centerline{ \psfig{file=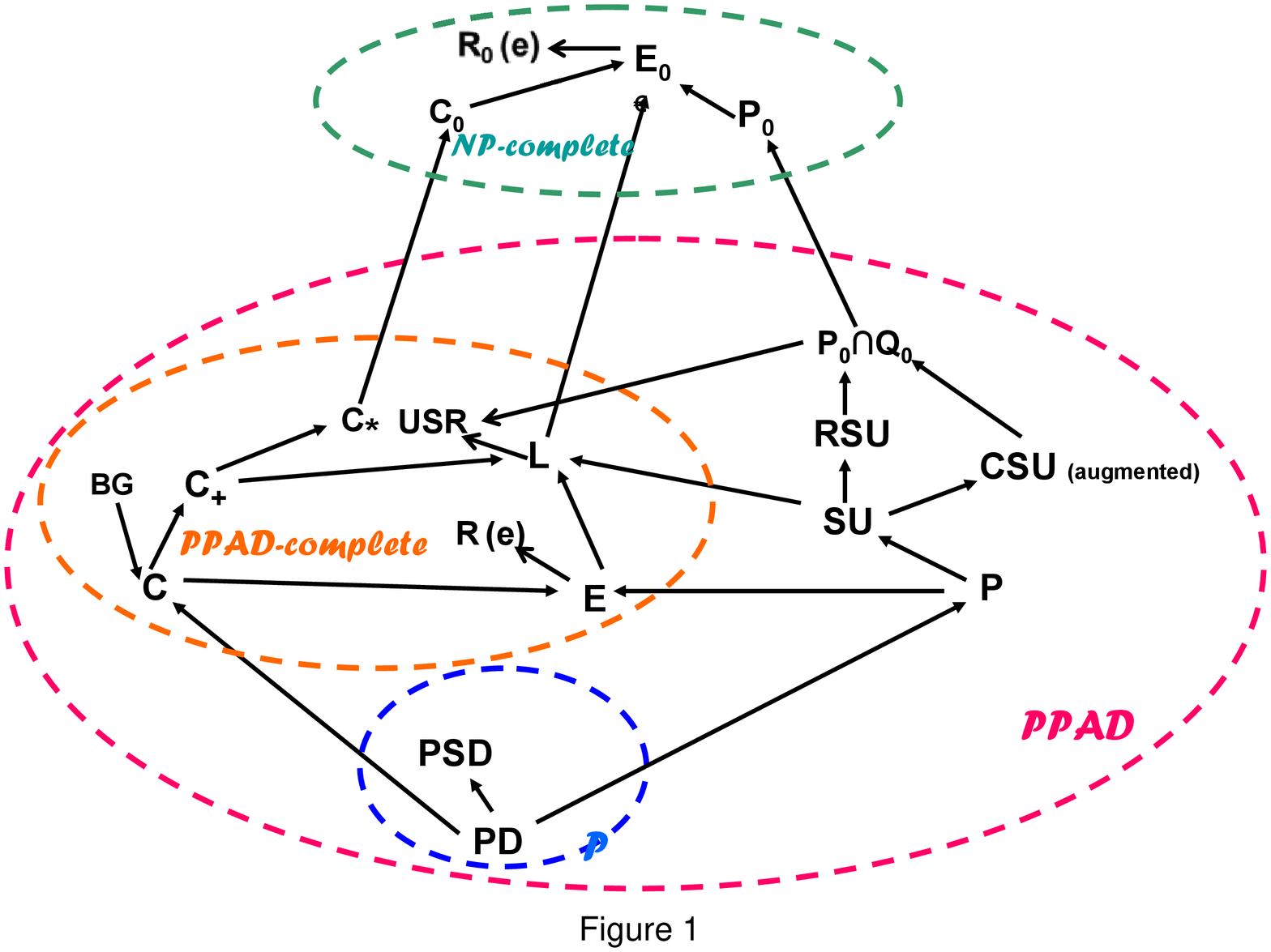  ,width=4.0in}}
       Next we show how Lemke($d$) $\ppad$-verified linear complementarity problems
       can be reduced simply and directly to a symmetric bimatrix game.
       To achieve this goal,
       we shall consider the following generic problem which we call {\it Lemke($d$)-\lcp(q,M)} and denote
        by $\llcp(d,q,M)$.

            \noindent
  $\llcp(d,q,M)$\;:\; Given $M \in \rmm,\; q \in \rm1,$ and $d \in \rmonepp$, find either
   $z \in \sol(q,M)$ or  $(\bar{z}_0,\bar{z},\bar{u}_0,\bar{u}) \in \sr(d,q,M)$.

     Obviously $\llcp(d,q,M)$ is verified to be in $\ppad$ by $G(d,q,M)$. In the following we shall show how it is possible
     to reduce any $\llcp(d,q,M)$, whenever $\lcp(d,M)$ is nondegenerate, to a symmetric bimatrix game. In addition,
     given $\bar{d}>0$, we shall show, by standard LP perturbation
     techniques, that our reduction works for all $d$ in a sufficiently small neighborhood of $\bar{d}$, and that the reduction works for all
     $d \in \ronepp$, except for a finite number of subsets of measure zero.
    We shall present our reduction in several steps, where we address the reduction of instances of the following problem:

      \noindent
  $\y-\llcp(d,q,M)$\;:\; Given $M \in \rmm,\; q \in \rm1,\; \d \in \rmonepp$ and a matrix class $\y$, find
 one of the following.

   (1) $z \in \sol(q,M)$,\; (2) $(\bar{z}_0,\bar{z},\bar{u}_0,\bar{u}) \in \sr(d,q,M)$,\; (3)    A certificate that $M \not \in \y$.

  We start by presenting in Section \ref{sec:re} a simple reduction which is applicable to
     $\r(e) - \llcp(e,q,M)$. The certificate that we obtain from the bimatrix game whenever $M \not \in \r(e)$
     is of the form $0 \neq \bar{u} \in \sol(e\bar{u}_0,M)$. So
      we get either a solution for $\lcp(q,M)$  or
    a secondary direction for $\elcp(e,q,M)$.  Note that at this stage we address only $e$ as a
 covering vector, and that we do not reach yet our goal as the reduction may produce secondary {\it directions} rather than secondary {\it rays}.
      In Section
     \ref{sec:r_0}, we  extend the previous reduction to handle $\r_0(e)-\llcp(e,q,M)$.
     The certificate that we obtain from the bimatrix game whenever $M \not \in \r_0(e)$
     is of the form $0 \neq \bar{u} \in \sol(e, M)$. So
      we get either a solution for $\lcp(q,M)$, a secondary ray for $\elcp(e,q,M)$, or
    a  type $1$ secondary direction for $\elcp(e,q,M)$.
     Next, in Section \ref{sec:nd}, we present a complete reduction of $\llcp(e,q,M)$ under the assumption
     that $\lcp(e,M)$ is nondegenerate if $\sol(e,M) \neq \{0\}$.
       Finally, in Section \ref{sec:ext}, we  extend the reductions in the previous sections to a general
     covering vector $d>0$ for which $\lcp(d,M)$ is nondegenerate if $\sol(d,M) \neq \{0\}$.
         \section{Reducing \boldmath{$\r(e)-\lcp(q,M)$} to a symmetric bimatrix game}
\label{sec:re}
   In this section we present a simple direct reduction of
   $\r(e)-\lcp(e,q,M)$. In particular, given $q,M$, we construct a symmetric bimatrix game
   whose symmetric Nash equilibrium points
   correspond one-to-one to either $\bar{z} \in \sol (q,M)$ or
    a certificate for $M \not \in \r(e)$ in the form of $0 \neq \bar{u} \in \sol(e\bar{u}_0,M)$ where
    $\bar{u}_0\in\{0,1\}$, so $(\bar{u}_0,\bar{u}) \in \sd(e,M)$.

 Given $\lcp(q,M)$ with $M \in \rmm, q \in \rm1$ and a covering vector $e$, we set $n=m+1$ and
 a symmetric bimatrix game whose cost matrix $C(q,M)$ is
  \begin{equation}
  \label{eq:C}
  C(q,M) \triangleq
   \left[
\begin{array}{cc}
M & q + e \\
 0 & 1 \\
\end{array}
\right].
 \end{equation}

Given $C(q,M)$ as above, we denote any symmetric equilibrium point
  $x \in \sne(C)$  as
  $x \triangleq
  \left(
\begin{array}{c}
y \\
 t \\
\end{array}
\right)$,
  where $y \in \rm1$ and $t \in \rone$.
 Given $\sne(C(q,M))$, we partition it to
  \[
    \sne_{+}(C(q,M)) \triangleq \left\{
  \left[
\begin{array}{c}
y \\
 t \\
\end{array}
\right]
 \in \sne(C(q,M))\;| \; t>0 \right\},
  \]
  and
   \[
    \sne_{0}(C(q,M)) \triangleq \left\{
  \left[
\begin{array}{c}
y \\
 t \\
\end{array}
\right]
 \in \sne(C(q,M))\;| \; t=0 \right\},
  \]
 In the next theorem we   establish a one-to-one correspondence  between the symmetric Nash equilibria
 of $SG(C(q,M))$ which use with positive probability the last column of $C(q,M)$, and the set of solutions to $\lcp(q,M)$.
 We follow this with a theorem that establishes a one-to-one correspondence  between the symmetric Nash equilibria
 of $G(C(q,M))$ which are not using the last column of $C(q,M)$ and the secondary directions of $\elcp(e,q,M)$.

  \vspace{-6mm}
  \theo
  \label{theo:lcp_nse+}
  \hphantom \newline \noindent
    \begin{description}
      \item[(i)]
    Given $
    \bar{x}= \left[
\begin{array}{c}
 \bar{y} \\
 \bar{t}\\
\end{array}
\right]
     \in \sne_{+}(C(q,M))$, let $\bar{z}= \bar{y}\frac{1}{\bar{t}}$. Then, $\bar{z} \in \sol(q,M)$.
    \item[(ii)]
     Given $\bar{z} \in \sol(q,M)$, let $\bar{t}=\frac{1}{e\supt \bar{z} +1},\; \bar{y}= \bar{z}\bar{t} $.
     Then,
    $\bar{x}=
    \left[
\begin{array}{c}
 \bar{y} \\
 \bar{t} \\
\end{array}
\right]
     \in \sne_{+}(C(q,M))$.
    \end{description}
    \etheo
 \noindent {\bf Proof.} \hspace{3mm} Throughout the proof  we denote $C(q,M)$ by $C$.
   \begin{description}
         \item[(i)]
 Since $\bar{t} >0$, then,  by (\ref{eq:ne_4}) (with $i=n$), $\bar{t}= \bar{x}\supt C \bar{x}$.
    Thus, by (\ref{eq:ne_1})--(\ref{eq:ne_2}),
 $M \bar{y}+ (q + e)\bar{t} \geq e \bar{t}, \; \bar{y} \geq 0$. Dividing by $\bar{t}$, we get
 $M \bar{z}  +q \geq 0,\; \bar{z} \geq 0.$
  In addition, by (\ref{eq:ne_4}) (for $i=1, \ldots, m$),
  $\bar{y}_i (M_{i \dt} \bar{y} + q_i\bar{t} +\bar{t}-\bar{t})=0$, so
  dividing by $\bar{t}^2$, substituting for $\bar{z}$, and summing over $m$,
  we get $0 = \sum_{i=1}^m\bar{z}_i (M_{\dt i} \bar{z} + q_i ) =\bar{z}\supt (M \bar{z} +   q)$.
      \item[(ii)]
    By (\ref{eq:lcp_1}), and setting $\bar{t}=\frac{1}{e\supt z +1},\; \bar{y}= \bar{z}\bar{t} $, we have
   \[
      \left[
\begin{array}{cc}
M & q + e \\
 0 & 1 \\
\end{array}
\right]
   \left[
\begin{array}{c}
 \bar{z}\bar{t}\\
 \bar{t} \\
\end{array}
\right]
\geq
 \left(
\begin{array}{c}
 e\\
 1\\
\end{array}
\right)
\left(
\begin{array}{c}
\bar{z}\bar{t}\\
\bar{t}\\
\end{array}
\right)
\geq
\left(
\begin{array}{c}
 0\\
 0\\
\end{array}
\right)
 \]
and obviously $(e\supt \bar{z})\bar{t} +\bar{t} =(e\supt \bar{z}+1)\bar{t}=1.$
 In addition,
\[
 \bar{x}\supt C \bar{x}= \bar{y}\supt (M \bar{y} + q\bar{t} +e\bar{t}) + \bar{t}^2 = \bar{t}^2 (\bar{z}\supt (M \bar{z} + q) +e) + 1)
\]

  Thus, since by (\ref{eq:lcp_2}), $\bar{z}\supt (M \bar{z} + q)=0$,
  $ \bar{x}\supt C \bar{x}=\bar{t}^2(\bar{z}\supt e +1)=\bar{t}$. Hence,
   $x=
   \left(
\begin{array}{c}
 \bar{y}\\
 \bar{t}\\
\end{array}
\right)$
   satisfies  (\ref{eq:ne_1})--(\ref{eq:ne_3}), and since
  $\bar{t}>0$, we have $\bar{x} \in \sne_{+}(C(q,M)). \hspace*{\fill} \qed$
  \end{description}
  \theo
  \label{theo:lcp_nse0}
     \hphantom \newline \noindent
      \begin{description}
      \item[(i)]
    If $
   \left[
\begin{array}{c}
\bar{y}\\
 0 \\
\end{array}
\right]
    \in \sne_{0}(C(q,M))$, then $\bar{y} \in \sol( e \tau,M)/\{0\}$ for some $\bar{\tau} \geq 0$.
    \item[(ii)]
    Let $ \bar{u}  \in \sol( e \bar{\tau},M)/\{0\}$ for some $\bar{\tau} \geq 0$.
    Then, setting $\bar{y} = \bar{u}  \frac{1}{e\supt \bar{u} }$,
    $
    \bar{x} =
   \left[
\begin{array}{c}
  \bar{y} \\
 0 \\
\end{array}
\right]
     \in
     \sne_{0}(C(q,M))$.
    \end{description}
    \etheo
 \noindent {\bf Proof.} \hspace{3mm} Throughout the proof  we denote $C(q,M)$ by $C$.
    \begin{description}
    \item[(i)]
  By \eqref{eq:ne_1}--\eqref{eq:ne_2},
  $M \bar{y}  \geq e(\bar{y} \supt M \bar{y} ),\; 0 \geq \bar{y} \supt M \bar{y}  ,\; \bar{y}  \geq 0,$
  and $e\supt \bar{y} =1$. Setting  $\bar{\tau} = - \bar{y} \supt M \bar{y} $, we get
  $M \bar{y}  +e\bar{\tau} \geq 0,\; 0 \neq \bar{y}  \geq 0$.
  Moreover, since $e\supt \bar{y}  =1$, we have (by \eqref{eq:ne_4}) that
 $\bar{y} \supt(M \bar{y}  +e {\bar{\tau}})=0$, concluding that $\bar{y}  \in \sol(e \bar{\tau},M)/\{0\}$
  for some $\bar{\tau} \geq 0.$
      \item[(ii)]
      Noticing that $\bar{u}  \neq 0$
      and by (\ref{eq:lcp_1})--(\ref{eq:lcp_2}),
      \[
      M \bar{y} \geq e(-\bar{\tau}),\;\; \bar{y}\supt(M\bar{y} + e(-\bar{\tau}))=0,\;\; \bar{y} \geq 0.
      \]

  Thus,
  $\left[
\begin{array}{c}
 \bar{y}\\
 0 \\
\end{array}
\right]$
  satisfy (\ref{eq:ne_1})--(\ref{eq:ne_2}).
  Noticing that $ e\supt \bar{u} =1$, completes the proof.
  $\hspace*{\fill} \qed$
     \end{description}
     Given $\lcp(q,M)$,  and combining Theorems \ref{theo:lcp_nse+} and \ref{theo:lcp_nse0},
 we can construct a symmetric bimatrix game where any
 symmetric Nash equilibrium point corresponds to either a solution for $\lcp(q,M)$, or a secondary direction
 for $\elcp(e,q,M)$.
 Specifically, given $q \not \geq 0$ and $M$, consider the symmetric bimatrix game whose cost matrix is
 $C(q,M)$.
  Let
  $
  \left[
\begin{array}{c}
 \bar{y} \\
  \bar{t} \\
\end{array}
\right] \in \sne(C(q,M))$, and let $\bar{\tau}$ be its expected cost.
 We then conclude that:
 \begin{description}
   \item[1] $\boldsymbol{\bar{t}>0}$. By Theorem \ref{theo:lcp_nse+}-(i),
   $\frac{1}{\bar{t}} \bar{y} \in \sol(q,M)$,
  \item[2] $\boldsymbol{\bar{t}=0}$. By Theorem \ref{theo:lcp_nse0}-(i),
  $\bar{y} \in \sol(e \bar{\tau} ,M)\setminus\{ 0 \}$ for
  some $\bar{\tau} \geq 0$,
  \begin{description}
    \item[1.1] $\boldsymbol{\bar{\tau}=0}$. Then, $\bar{y} \in \sol(0,M)$ with $e\supt \bar{y} =1$,
     so $\bar{y} \in \sd_0(d,q,M)$,
    \item[1.2] $\boldsymbol{\bar{\tau}>0}$,
  Then,
   $\bar{y} \frac{1}{\bar{\tau}} \in \sol(e,M)$, so $\bar{y}\frac{1}{\bar{\tau}} \in \sd_1(e,q,M)$.
      \end{description}
    \end{description}
  So the symmetric bimatrix game $SG(C)$ generates either a solution for $\lcp(q,M)$ or a secondary
  direction for $\elcp(e,q,M)$.

    \noindent {\bf Remarks} \hspace{1mm}
    \begin{enumerate}
    \item
    Note that $\r(e)-\llcp(e,q,M)$ is also reduced to $SG(C(q,M))$.
    \item
    The class of all matrices $M$ for which $\lcp(q,M)$ is guaranteed to have a solution for all $q$ is called $\q$. The largest known class
    $\Y$ which is contained in $\q$ and for which it is known that $\Y-LCP(q,M)$ is Lemke(e)-resolvable, is $\r(e)$.
    \item
    Since $\p,\c \subset \e \subset \r(e)$, the reduction is applicable to $\y-\lcp(q,M)$, where $\Y$ is $\p,\c$ or $\e$.
    Note that $\lcp(q,M)$ has a unique solution for all $q$ if and only if $M \in \p$, and that
    $\lcp(q,M)$ has a unique solution for all $q \geq 0$ if and only if $M \in \e$.
   \end{enumerate}
          \section{Reducing \boldmath{$\r_0(e) -\llcp(e,q,M)$} to a symmetric bimatrix game}
\label{sec:r_0}
    In this section we consider $\r_0(e)-\llcp(e,q,M)$ which brings us closer to achieving our goal of
    reducing any $\llcp(e,q,M)$ to a symmetric bimatrix game.
    In particular, given $q,M$, we construct a symmetric bimatrix game
   whose symmetric Nash equilibrium points
   correspond one-to-one to either the solutions of $\lcp(q,M)$, the  secondary rays
   of $\elcp(e,q,M)$, or
    certificates for $M \not \in \r_0(e)$ in the form of $0 \neq \bar{u} \in \sol(e,M)$
    which correspond to a type 1 secondary directions of  $\elcp(e,q,M)$.

    For that purpose we
     introduce
    the  {\it augmented} problem $\lcp(\tilde{q},\tilde{M})$ associated with  $\lcp(q,M)$,
   where
      \[\tilde{M}=
     \left[ \begin{array}{rr}
      1   & -e \supt \\
      e   &   M
\end{array} \right],\;
\tilde{q}=
  \left[ \begin{array}{c}
   \beta\\
   q \\
\end{array} \right],
\]
\noindent
 and $\beta > e\supt \bar{z} $ for any  vertex (not necessarily feasible)
 $(\bar{z}_0, \bar{z})$ of $\elcp(e,q,M)$.

  \noindent {\bf Remarks} \hspace{1mm}
  \begin{enumerate}
  \item
  It is a standard result in LP theory that if the entries in $q,M$ are rational then $\beta$
  is of size polynomial in the size of
  $\lcp(q,M)$, and that $\beta$  can be computed in time polynomial in the size of
  $\lcp(q,M)$.
  \item
  Augmented LCP systems where  $\tilde{M}_{11}$ is equal to $0$ (see \cite{cps92}) or $-1$ \cite{t73} are used in the LCP literature to
  eliminate  secondary rays. Such augmentations do not work in our case since the reduction of
  $\lcp(\tilde{q},\tilde{M})$ to a symmetric bimatrix game would yield a pure Nash equilibrium (using with probability 1 the strategy corresponding to the
   first column of $\tilde{M}$) which yields no information about the solution
  (or lack thereof) of the original $\lcp(q,M)$. To avoid this possibility,
  we need $\tilde{M}_{11}>0$, hence the choice of $1$.
  \end{enumerate}

   In the following theorem we establish the relationship between $\lcp(q,M)$ and  $\lcp(\tilde{q},\tilde{M})$.
   \vspace{-6mm}
  \theo
\label{theo:aug_lcp}
   \hphantom \newline \noindent
   \begin{description}
    \item[(i)]
    $ \left[ \begin{array}{c}
 0 \\
   \bar{z}
\end{array} \right]
     \in \sol(\tilde{q},\tilde{M})$
     if and only if
   $\bar{z} \in \sol(q,M)$.
  \item[(ii)]
   If
   $
   \left[ \begin{array}{c}
   \tilde{z}_0 \\
     \tilde{z}
\end{array}
 \right]
\in \sol(\tilde{q},\tilde{M})$ where $\tilde{z}_0 >0$,
   then there exists $( \bar{z}_0,\bar{z},\bar{u}_0,\bar{u})\in \sr(e,q,M)$ and $\bar{\lambda}>0$ such that
   \[
   \left[ \begin{array}{c}
   \tilde{z}_0 \\
     \tilde{z}
\end{array}
 \right]=
 \left[ \begin{array}{c}
   \bar{z}_0 \\
     \bar{z}
\end{array}
 \right]
 +\bar{\lambda}
 \left[ \begin{array}{c}
   \bar{u}_0 \\
     \bar{u}
\end{array}
 \right].
 \]
    \item[(iii)]
    Let  $( \bar{z}_0,\bar{z},\bar{u}_0,\bar{u})\in \sr(e,q,M)$. Then there exists
    $\bar{\lambda}$ such that
    $
    \left[ \begin{array}{c}
  \bar{z}_0 \\
   \bar{z}
\end{array} \right]
 +
 \bar{\lambda}
   \left[ \begin{array}{c}
  \bar{u}_0 \\
   \bar{u}
\end{array} \right]
    \in \sol(\tilde{q},\tilde{M}).$
   \end{description}
\etheo
 \noindent {\bf Proof.} \hspace{3mm}
    \begin{description}
    \item[(i)]
    The `only if' direction is obviously true. The `if' direction is  true because of the nondegeneracy assumption
    (so $\tilde{z}$ is a a vertex of $\lcp(q,M)$) and by the definition of $\beta$.
      \item[(ii)]
      Let
   $\left[ \begin{array}{c}
  \tilde{z}_0 \\
  \tilde{z}
\end{array} \right] \in \sol(\tilde{q},\tilde{M})$ where $\tilde{z}_0>0$.
    Then,
  \[
   q+  e \tilde{z}_0 + M \tilde{z}  \geq 0,\; \tilde{z} \geq 0,\; \tilde{z}_0  \geq 0,\;
    \tilde{z}\supt (q+  e \tilde{z}_0 + M \tilde{z} )=0, \; \tilde{z}_0 (\beta + \tilde{z}_0 - e\supt \tilde{z})=0,
\]
    which implies that $(\tilde{z}_0,\tilde{z}) \in \elcp(e,q,M)$ and (since $\tilde{z}_0,>0$)
    $e\supt \tilde{z} = \beta + \tilde{z}_0$.
    However, by the definition
    of $\beta$,
      $(\tilde{z}_0,\tilde{z})$ must be a point on a secondary ray of $\elcp(e,q,M)$. That is, there exists
      $\bar{\lambda}>0$ and
      $(\bar{z}_0, \bar{z}, \bar{u}_0, \bar{u}) \in \sr(e,q,M)$  such that
               $
                 \left[ \begin{array}{c}
  \tilde{z}_0 \\
   \tilde{z}
\end{array} \right]
=
    \left[ \begin{array}{c}
  \bar{z}_0 \\
   \bar{z}
\end{array} \right]
 +
 \bar{\lambda}
   \left[ \begin{array}{c}
   \bar{u}_0 \\
   \bar{u}
\end{array} \right]$. 
       \item[(iii)]
       By the definition of $\sr(e,q,M)$, we have that
         for all $\lambda \geq 0$,
           $
   \left[ \begin{array}{c}
  \bar{z}_0 \\
   \bar{z}
\end{array} \right]
 +
 \lambda
   \left[ \begin{array}{c}
   \bar{u}_0 \\
   \bar{u}
\end{array} \right]$
  satisfies all the constraints of $\lcp(\tilde{q},\tilde{M})$ except possibly for the last constraint.
        However,  since
        $-  \bar{z}_0  + e\supt \bar{z} < \beta$, $\bar{u} \neq 0$, $\bar{u}_0  \in \{0,1\}$,
        and $-\bar{u}_0 + e \supt \bar{u}>0$,
        setting $\bar{\lambda}=\frac{\beta + \bar{z} _0- e\supt \bar{z}}{-\bar{u}_0 + e\supt \bar{u}}$
    yields $- (\bar{z}_0 + \bar{\lambda} \bar{u}_0) + e\supt (\bar{z} + \bar{\lambda} \bar{u}) =\beta$ which,
  considering that $\bar{\lambda} >0$, completes
      the proof.
  $\hspace*{\fill} \qed$
     \end{description}

      \noindent {\bf Remark} \hspace{1mm}
      The extraction of either a solution or a secondary ray from a secondary direction as described in
      the proof of Theorem \ref{theo:aug_lcp}-(ii) can be done by standard LP technique that can be executed in strongly polynomial time
      (that is, the required number of elementary calculations such as additions, multiplications, divisions and comparisons is bounded
      above by a polynomial function of $m$).

       Next, we show that $\tilde{M} \not \in \r(e)$ implies that $M \not \in \rzero(e)$, which allows us to apply the reduction of
     the previous section to the augmented problem.
   \theo
   \label{theo:R_0}
   If
   $ 0 \neq
\left[ \begin{array}{c}
   \bar{u}_0\\
  \bar{u}
\end{array} \right]
 \in \sol(e \tau,\tilde{M})
$
   for some $\tau \geq 0$, then $\bar{u}_0+\bar{\tau} >0$ and $\frac{1}{\bar{u}_0+\bar{\tau}} \bar{u} \in \sol(e,M)$.
 \etheo
\noindent {\bf Proof.} \hspace{3mm}
 By the premise of the theorem
 there exists
 $ \bar{u}_0 \geq 0, \;\bar{u} \geq 0,\; \bar{u}_0 + e\supt \bar{u}>0$,
such that
  \[
  \left[ \begin{array}{cc}
   1 & -e\supt \\
  e  & M \\
\end{array} \right]
\left[ \begin{array}{c}
   \bar{u}_0\\
   \bar{u}
\end{array} \right] +
  \left[ \begin{array}{c}
  1 \\
   e\\
\end{array} \right]
\bar{\tau}
\geq
 \left[ \begin{array}{c}
   0\\
   0
\end{array} \right],\;
\mbox{ and }\]
\[
[\bar{u}_0 \; \bar{u}\supt ]
 \left(
 \left[ \begin{array}{cc}
  1 &   -e\supt \\
  e  & M \\
\end{array} \right]
\left[ \begin{array}{c}
   \bar{u}_0\\
   \bar{u}
\end{array} \right] +
  \left[ \begin{array}{c}
   1\\
   e\\
\end{array} \right]
\bar{\tau}
\right)
=
 \left[ \begin{array}{c}
   0\\
   0
\end{array} \right].
  \]
  Thus,
  \begin{subequations}
    \label{eq:aug_sys}
    \begin{eqnarray}
     \label{eq:aug_sys_1}
   \bar{u}_0  -e\supt \bar{u} +  \bar{\tau} \geq 0,
 \hspace{47mm}\\
  \label{eq:aug_sys_2}
   M \bar{u} + e(\bar{u}_0 +  \bar{\tau}) \geq 0,\; \bar{u} \geq 0,
   \hspace{30mm}\\
    \label{eq:aug_sys_3}
   \bar{u}\supt(M \bar{u} +e(\bar{u}_0 +  \bar{\tau}))=0.
   \hspace{35.2mm}
  \end{eqnarray}
  \end{subequations}
  By \eqref{eq:aug_sys_1} we have that  $\bar{u}_0 + \bar{\tau} >0$ (as otherwise $\bar{u}=0, \bar{u}_0=0$ contrary to the assumption). Thus,
  from \eqref{eq:aug_sys_2} and \eqref{eq:aug_sys_3} we have that $\frac{1}{\bar{u}_0+\bar{\tau}}\bar{u} \in \sol(e,M). \hspace*{\fill} \qed $

\medskip

Combining Theorems \ref{theo:aug_lcp} and \ref{theo:R_0}, and recalling the definition of the class
$\r_0(e)$, we get that, given $\lcp(q,M)$, we can construct a symmetric bimatrix game  where any
 symmetric Nash equilibrium point corresponds to a solution of $\lcp(q,M)$, a  secondary ray
 of $\elcp(e,q,M)$ or a type 1 direction of $\elcp(e,q,M)$.

 Specifically, consider the symmetric bimatrix game whose cost matrix is
$
 C(\tilde{q}, \tilde{M})
 =
 \left[
\begin{array}{ccc}
 1 & -e\supt &  -\beta +1 \\
 e& M& q + e \\
0 &  0 & 1 \\
\end{array}
\right]$.
  Let
  $
  \left[
\begin{array}{c}
 \bar{z}_0 \\
  \bar{z} \\
  \bar{t} \\
\end{array}
\right] \in \sne(C(\tilde{q},\tilde{M}))$, and let $\bar{\tau}$ be its expected cost.
 We then conclude that:
 \begin{description}

  \item[1] $\boldsymbol{\bar{t}>0}$.
  By Theorem \ref{theo:lcp_nse+},
  $
  \left[
\begin{array}{c}
   \bar{z}_0\\
 \bar{z} \\
\end{array}
\right]
 \frac{1}{\bar{t}}
 \in \sol(\tilde{q},\tilde{M}).$
 \begin{description}
    \item[1.1] $\boldsymbol{\bar{z}_0=0}$. Then (by Theorem \ref{theo:aug_lcp}-(i)), $\bar{z}\frac{1}{\bar{t}} \in \sol(q,M)$ (a solution to the original problem).
    \item[1.2] $\boldsymbol{\bar{z}_0>0}$.
  Then (by Theorem \ref{theo:aug_lcp}-(ii) and the remark following its proof),  we can obtain (in strongly polynomial time)
   $(\bar{z}_0,\bar{z},\bar{u}_0,\bar{u}) \in \sr(e,q,M)$ (a secondary ray of $\elcp(e,q,M)$).
      \end{description}
      \item[2] $\boldsymbol{\bar{t}=0}$. By Theorem \ref{theo:lcp_nse0}.
        $
   0
 \neq
\left[ \begin{array}{c}
   \bar{u}_0\\
  \bar{u}
\end{array} \right]
 \in \sol(e \bar{\tau},\tilde{M})
$
  where $\bar{\tau}\geq 0$. Thus, by Theorem  \ref{theo:R_0},

  \noindent
    $ 0 \neq \bar{u} \frac{1}{\bar{u}_0+\bar{\tau}} \in \sol(e,M)$
    (so
    $
\left[ \begin{array}{c}
  1 \\
  \bar{u} \frac{1}{\bar{u}_0+\bar{\tau}}
\end{array} \right] \in \sd_1(e,q,M)
$
 (a  type 1 secondary direction of $\elcp(e,q,M)$).
  \end{description}

           \section{Handling nondegenerate type 1 secondary directions}
\label{sec:nd}
   In this section we show that if $0 \neq \bar{u} \in \sol(e,M)$ is nondegenerate, then we can
    compute, in strongly polynomial time, ($\bar{z}_0,\bar{z}$), a vertex of $\elcp(q,M)$ such that
    if $\bar{z}_0 >0$ then $(\bar{z}_0,\bar{z},1,\bar{u}) \in \sr_1(e,q,M)$ (that is, a type 1 secondary ray).

    Let $\bar{u}$ be a non-zero, nondegenerate solution for $\lcp(e,M)$.
    Setting $\bar{v}=M\bar{u}+e$,  let
    $\alpha=\{i \;|\; \bar{u}_i >0\}$ and $\bar{\alpha}=\{i \;|\; \bar{v}_i >0\}$. Note that by the nondegeneracy
    assumption $\alpha \cup \bar{\alpha}=\{1, \ldots,m\}$ and $ M_{\alpha\alpha}$ is nonsingular.
    Now, we set
    \[
    \left[ \begin{array}{c}
   \hat{z}_\alpha\\
  \hat{w}_{\bar{\alpha}}
\end{array} \right]
=
 \left[ \begin{array}{cc}
   M_{\alpha \alpha} & 0\\
   M_{\bar{\alpha}\alpha} & I
\end{array} \right]^{-1}
\left[ \begin{array}{c}
   q_\alpha\\
  q_{\bar{\alpha}}
\end{array} \right],
\;
  \left[ \begin{array}{c}
   \hat{z}_{\bar{\alpha}}\\
  \hat{w}_{\alpha}
\end{array} \right]
=
  \left[ \begin{array}{c}
  0\\
 0
\end{array} \right],
    \]
     which results in
     \[\hat{w}= q + M\hat{z},\; \hat{z}\supt \hat{w}=0,\; \bar{u}\supt(q + M\hat{z}),\;\hat{z}\supt(M\bar{u}).
      \]
      Thus, if $\hat{z},\hat{w} \geq 0$ then $\hat{z} \in \sol(q,M)$. Otherwise, since $\bar{u}_\alpha >0$ and
      $\bar{v}_{\bar{\alpha}} >0$, for sufficiently large $\lambda >0$ we have that
        \[
    \left[ \begin{array}{c}
   \hat{z}_\alpha\\
  \hat{w}_{\bar{\alpha}}
\end{array} \right]
    +
   \lambda  \left[ \begin{array}{c}
   \bar{u}_\alpha\\
  \bar{v}_{\bar{\alpha}}
\end{array} \right]
\geq
\left[ \begin{array}{c}
   0\\
  0
\end{array} \right]
\]
  Letting $\bar{\lambda}$ be the smallest $\lambda$ satisfying the inequality above, and setting
  $\bar{z}_0=\bar{\lambda}, \bar{z}=\hat{z} + \bar{\lambda} \bar{u}$ we get that
  $(\bar{z}_0,\bar{z},1 , \bar{u}) \in \sr_1(e,q,M)$. Note that constructing ($\bar{z}_0,\bar{z}$) whenever a nondegenerate
 \section{Extensions}
 \label{sec:ext}
   In this section we show how to extend our reductions whenever a general positive covering vector is used rather than $e$.
   The key to the results in this section is the following proposition.
   \prop
   \label{prop:M(d)}
  \hphantom \newline \noindent
  Given $d \in \rmonepp$, let $D$ be the diagonal matrix with $D_{ii}=d_i$. Then,
    $(\bar{z}_0,\bar{z}) \in \elcp(d,q,M)$ if and only if $(\bar{z}_0,D\bar{z}) \in \elcp(e,D^{-1}q,D^{-1}MD^{-1})$.
   \eprop
     \noindent {\bf Proof.} \hspace{3mm}
      The proposition is easily verified by observing that

     \noindent
      $q + d z_0  + Mz  \geq 0,\; z \geq  0\;$ if and only if $\;D^{-1}(q + d z_0 + MD^{-1}Dz ) \geq 0,\; Dz \geq  0$,

     \noindent
      and $ z \supt(q + d z_0 + Mz )=0 \;$ if and only if $z \supt D D^{-1}(q + d z_0 + MD^{-1}Dz )=0. \;\hspace*{\fill} \qed$

    \coro
   \label{coro:M(d)}
   Given $d \in \rmonepp$, let $D$ be the diagonal matrix with $D_{ii}=d_i$. Then,
  \begin{description}
  \item[(i)]
 $\bar{z} \in \lcp(q,M)$ if and only if $D\bar{z} \in \lcp(D^{-1}q,D^{-1}MD^{-1})$.
   \item[(ii)]
   $M \in \r(d)$ if and only if $D^{-1} M D^{-1} \in \r(e)$.
 \item[(iii)]
   $M \in \rzero(d)$ if and only if $D^{-1} M D^{-1}\in \rzero(e)$.
   \end{description}
   \ecoro
   \noindent {\bf Proof.} \hspace{3mm}
      \begin{description}
  \item[(i)]
  Results from Proposition \ref{prop:M(d)} by considering $\bar{z}_0=0$.
   \item[(ii)-(ii)]
   Result directly from (i)  and the definitions of $\r(d)$ and $\rzero(d)$.
   \end{description}

   By Proposition \ref{prop:M(d)} and Corollary  \ref{coro:M(d)} it can be readily verified that the results
   of Section \ref{sec:re} can be extended to $\r(d) - \lcp(q,M)$ by considering  $\r(e) - \lcp(D^{-1}q,D^{-1}MD^{-1})$.
   Similarly, the results of Section \ref{sec:r_0} can be extended  to $\rzero(d) - \llcp(d,q,M)$ by considering
    $\rzero(e) - \llcp(e,D^{-1}q,D^{-1}MD^{-1})$. Finally, given that $\lcp(d,M)$ is nondegenerate we can (by
    Proposition \ref{prop:M(d)}, Corollary  \ref{coro:M(d)}, and the results of sections \ref{sec:r_0} and \ref{sec:nd})
    reduce any $\llcp(d,q,M)$ where $\lcp(d,M)$ is nondegenerate
    to a bimatrix game whose cost matrix is
    \[
 C=\left[
\begin{array}{ccc}
 1 & -e\supt &  -\beta\prod_{i=1}^{m} d_i +1 \\
 e& D^{-1}MD^{-1}   & D^{-1}q + e \\
0 &  0 & 1 \\
\end{array}
\right].
\]
 Let
  $
  \left[
\begin{array}{c}
 \bar{z}_0 \\
  \bar{z} \\
  \bar{t} \\
\end{array}
\right] \in \sne(C)$. Based on Proposition \ref{prop:M(d)} and Corollary \ref{coro:M(d)} we replace $\bar{z}$
with $D^{-1}\bar{z}$ and proceed to recover a `solution' to $\llcp(d.q.M)$ by following the steps prescribed
in sections \ref{sec:r_0} and \ref{sec:nd}.
   \section{Concluding remarks}
 \label{sec:remarks}
\begin{enumerate}
  \item
  The main result of this paper is that  for almost any given $d\in \rmonepp,q \in \rmone, M \in \rmm$,  it is possible to effectively set up a symmetric bimatrix game whose Nash equilibria correspond one-to-one to all the endpoints (excluding the one corresponding to the primary ray)
  of the directed graph associated with $\lcp(q,M)$ and  the Lemke method with a covering vector $d$. The only condition is that any Nash equilibrium  corresponding to a type $1$ secondary direction (that is a solution for $\lcp(d,M)$),  has to be nondegenerate. Note that if this is not the case, we can perturb (by standard LP techniques) $d$ to a $\hat{d}$ which, when used as the covering vector, will guarantee that  the reduction will work. This observation also means that for  any given $M,q$ the reduction is workable for all covering vectors
  $d \in \rmonepp$, with the exception of a finite number of sets of measure $0$.
  \item
  As a consequence of the main result as specified above, the reduction will resolve any $\lcp(q,M)$  which  is
  Lemke($d$) $\ppad$-verified. Note that all the major matrix classes of $M$ which are known to be Lemke($d$) resolvable for all $q$
  are actually known to be Lemke($d$) $\ppad$-verified (and typically for all $d \in \rmonepp$). These classes (which are all  subsets of
  $\usr(d)$) and their relationships are depicted in Figure $1$.
  \item
  The direct reductions which are presented in this paper highlight the importance of the problem of 2-NASH within mathematical programming. In a sense, we show that as any LP can be directly reduced to zero-sum game (see \cite{d51} and \cite{a12}), it is analogously possible to directly reduce any $\lcp(q,M)$ which is Lemke($d$)  $\ppad$-verified to a 2-NASH problem, thus showing that many of the results regarding
  such problems are relevant to LCP theory.
 \item
  The reductions in sections \ref{sec:re}, \ref{sec:r_0}, and \ref{sec:nd} are simple and easy to execute. Thus, any algorithm that is applicable to bimatrix games can be directly used to solve instances of $\rsu(d)-\lcp(q,M)$ and $\llcp(d,q,M)$. Also, considering that the proposed reductions are bijections, algorithms  with a variety of goals, such as enumerating all, or specific subsets of Nash equilibrium points can be applied for similar goals regarding the solutions of  linear complementarity problems for which our reductions are applicable. It should be noted that there is a vast literature covering the subjects of computing and enumerating Nash equilibria of bimatrix games (see e.g. the surveys in \cite{vs02},\cite{vs07} and the papers introduced in \cite{vs10}).
 \item
  Over the years  several refinements of Nash equilibrium have been introduced.
   In particular, some results  regarding the existence and computation of these refinements have been established. In
   \cite{mt98} some of these refinements are generalized to LCPs. The reduction of $\lcp(q,M)$ which are
   Lemke($d$) $\ppad$-verified (e.g. where $M \in \usr(d)$) to  symmetric NASH-2, provides us with a tool to investigate analogous questions with respect to the generalized refinements to such LCPs. For example, in a forthcoming paper, we demonstrate such an  analysis by proving that any $\lcp(q,M)$ with $M \in \r(e)$ has a proper solution.
    As a corollary of this analysis we
   prove that the (unique) solution of $\lcp(q,M)$ where $M \in \p$ is proper and thus settle a conjecture  posed in \cite{mt98}
   (where it is proved  for $2 \times 2$ matrices).
   \item
   The simple reductions proposed in this paper allow us, whenever applicable, to potentially gain additional insight into the nature
   of models represented by these LCPs. This seems to be especially useful for economic models such as market equilibrium.
  \end{enumerate}

\end{document}